\documentclass[12pt,bibliography]{article}
\usepackage{hyperref}
\usepackage{epsfig}
\usepackage{amsfonts} 
\usepackage{amsmath} 
\usepackage{color}

\usepackage[T1]{fontenc}
\usepackage[utf8]{inputenc}

\def\e{\eta}

\def\Or[#1]{{\text{O}}\left({#1}\right)}
\def\dotl[#1,#2]{\left\langle #1, #2 \right\rangle}
\def\dotlb[#1,#2]{[ #1, #2 ]}
\def\dotp[#1,#2]{(#1) \cdot (#2)}
\def\aff[#1,#2]{\hat{#1}(#2)}
\def\n4sym{{\cal N}=4 SYM}
\def\>{\rangle}
\def\<{\langle}

\def\weight[#1,#2,#3]{\{(#1),#2,#3\}}
\def\ads[#1]{$\text{AdS}_{#1}$}

\newcommand{\ba}{\begin{eqnarray}}
\newcommand{\ea}{\end{eqnarray}}

\title{Conformal Blocks in Mellin Space}
 
\newcommand{\be}{\begin{equation}}
\newcommand{\ee}{\end{equation}}  
\newcommand{\bi}{\begin{itemize}}
\newcommand{\ei}{\end{itemize}}

\newcommand{\Ocal}{{\mathcal O}}

\newcommand{\aslash}[1]{\,\,{\raise.15ex\hbox{/}\mkern-12mu #1}}
\newcommand{\bslash}[1]{\,\,{\raise.15ex\hbox{/}\mkern-9mu #1}}

\renewcommand{\tilde}{\widetilde}
\renewcommand{\hat}{\widehat}

\newcommand\lrpar{\raise .8ex\hbox{$^\leftrightarrow$} \hspace{-9pt}
\partial}
\newcommand\lpar{\raise .8ex\hbox{$^\leftarrow$} \hspace{-9pt}
\partial}
\newcommand\rpar{\raise .8ex\hbox{$^\rightarrow$} \hspace{-9pt}
\partial}
\newcommand\lrd{\raise .8ex\hbox{$^\leftrightarrow$} \hspace{-9pt}
\nabla}

\newcommand{\gsim}{\lower.7ex\hbox{$\;\stackrel{\textstyle>}{\sim}\;$}}
\newcommand{\lsim}{\lower.7ex\hbox{$\;\stackrel{\textstyle<}{\sim}\;$}}

 \let\b=\beta  \let\d=\delta \let\e=\epsilon
\let\z=\zeta    
\let\l=\lambda  \let\n=\nu  
\let\s=\sigma     
  \let\D=\Delta

\renewcommand{\ba}{\begin{eqnarray}}
\renewcommand{\ea}{\end{eqnarray}}
\newcommand{\bea}{\begin{eqnarray}}
\newcommand{\eea}{\end{eqnarray}}

\begin{document}

\begin{titlepage}

\begin{center}
\vspace{1cm}

{\Large \bf  Conformal symmetry \\
of the   
critical 3D Ising model inside a sphere }

\vspace{0.8cm}

{\bf  Catarina Cosme$^{1\,2}$, J. M. Viana Parente Lopes$^{1\,3}$, Jo\~ao Penedones$^{1\,2}$}

\vspace{.5cm}

{\it  $^1$ 
Departamento de Física e Astronomia\\
Faculdade de Ciências da Universidade do Porto\\
Rua do Campo Alegre 687, 4169–007 Porto, Portugal}

{\it  $^2$ Centro de Física do Porto\\ 
Faculdade de Ciências da Universidade do Porto\\
Rua do Campo Alegre 687, 4169–007 Porto, Portugal}

{\it $^3$  Centro de Física, Universidade do Minho,\\
 P-4710-057, Braga, Portugal}

\end{center}
\vspace{1cm}

\begin{abstract}
We perform Monte-Carlo simulations of the three-dimensional Ising model at the critical temperature and zero magnetic field. 
We simulate the system in a ball with free boundary conditions on the two dimensional spherical boundary. Our results for one and two point functions in this geometry are consistent with the predictions from the conjectured conformal symmetry of the critical Ising model.
\end{abstract}

\bigskip
\bigskip

\end{titlepage}

\tableofcontents
\section{Introduction}

It is a long standing conjecture that continuous phase transitions are described by conformal invariant field theories \cite{Polyakov:1970xd}.
Under reasonable assumptions  this conjecture has been proven in two dimensions \cite{Zamolodchikov:1986gt, Polchinski:1987dy} and, recently, in four dimensions \cite{arXiv:1309.2921, Dymarsky:2014zja}.
A general proof in three dimensions has not yet been found. However, assuming the validity of this conjecture it has been possible to formulate conformal bootstrap equations and find approximate solutions that predict the Ising critical exponents with high accuracy \cite{Rattazzi:2008pe, arXiv:1203.6064, arXiv:1403.4545, arXiv:1406.4858, arXiv:1502.02033}. This success can also be viewed as strong evidence for  conformal invariance of the 3D Ising model at the critical temperature.

Conformal invariance of the critical 3D Ising model can also be tested directly with lattice Monte-Carlo simulations.
In particular, in this work, 
we test the predictions of conformal symmetry for the critical Ising model in a ball with free boundary conditions on the two dimensional spherical boundary.
In two dimensions, the analogous geometry (disk) was analyzed in \cite{PhysRevE.67.036107}.
There have been other Monte-Carlo studies of conformal invariance in the 3D Ising model.
Using the standard cubic lattice hamiltonian, 
\cite{arXiv:1304.4110} showed that some two point functions in the presence of a line defect have the  functional form predicted by conformal invariance.
In  \cite{PhysRevE.67.066116}, the  authors used an anisotropic hamiltonian (with a continuous direction) to simulate the 3D Ising model  in several cylindrical geometries and measured correlation functions compatible with conformal invariance.
See also \cite{arXiv:1212.6190, arXiv:1407.7597} for an alternative implementation of a 3D cylindrical geometry.

\section{Ising model and Conformal Field Theory}
 
The Ising hamiltonian is
\be
H[\{s\}]=- \sum_{<x,y>} s(x) s(y)
\label{IsingHam}
\ee
where the local spin variables can take two values $s(x)=\pm 1$ and the sum is over nearest neighbours in a cubic lattice.
We are interested in correlation functions of local operators at the critical temperature
\be
\langle O_1(x_1) \dots O_n(x_n) \rangle = 
\frac{1}{Z} \sum_{\{s\}} e^{-\beta_c H[\{s\}]} O_1(x_1) \dots O_n(x_n)\ ,
\ee
where the partition function is
\be
Z= \sum_{\{s\}} e^{-\beta_c H[\{s\}]} \ .
\label{IsingZ}
\ee
The local operators in the Ising model can be classified by their quantum numbers with respect to the $\mathbb{Z}_2$ spin-flip symmetry and the 
point group symmetry of the cubic lattice.
The simplest local lattice operator that is invariant under the lattice symmetries that preserve the point $x$ and is odd under spin flip is the local spin field $s(x)$. In the sector of  operators invariant under the spin-flip symmetry and the lattice symmetries the simplest local operators are the identity $I$ and the energy density
\be
e(x)=\frac{1}{6} \sum_{\d} s(x)s(x+\d)
\ee
where $x+\d$ runs over the 6 nearest neighbours of $x$.

At the critical temperature, the Ising model has infinite correlation length and its correlation functions decay as power laws of the distances $|x_i-x_j|$.
The conjecture we want to test is that 
we can  define a Conformal Field Theory (CFT) that describes the correlators for $|x_i-x_j|$ much greater than the lattice spacing $a$.
In this continuum limit, the local lattice operators can be written in terms of the operators of the Ising CFT that have the same symmetry properties. In particular, the spin field can be expanded in terms of the CFT scalar operators $\s,\, \s',\,\dots $which are odd under spin-flip
\footnote{In general this expansion also includes descendant scalar operators, which we did not write to avoid cluttering. 
In particular, the operator $\partial^2 \sigma$ is also present and has lower dimension than $\sigma'$.}
\begin{align}
s(x)&= b_{s\s}\, a^{\D_\s} \s(x) + b_{s\s'} \,a^{\D_{\s'}} \s'(x) + \dots\ ,\ \  \ 
&\D_\s &< \D_{\s'} < \dots  
\label{ssigma}
\end{align} 
where $\D_{\mathcal{O}}$ is the scaling dimension of the operator $\mathcal{O}$ and the $b$'s are dimensionless constants that depend on the normalization convention of the CFT operators.
Similarly, the local energy density operator can be written as 
\begin{align} 
e(x)&= b_{eI} \,I  + b_{e\e}\, a^{\D_\e} \e(x) + b_{e\e'} \,a^{\D_{\e'}} \e'(x) + \dots\ ,\ \ \ \ \ \  
&\D_\e &< \D_{\e'} < \dots
\label{eIeps}
\end{align} 
where $I$ is the identity operator and $\e,\,\e',\dots$ are the lowest dimension scalar primary operators in the $\mathbb{Z}_2$-even sector.
The best estimates for these scaling dimensions are \cite{ arXiv:1502.02033,   arXiv:1406.4858, arXiv:1403.4545, arXiv:1203.6064}
\be
\D_\s = 0.518151(6)\ ,\qquad
\D_{\s'}\gsim 4.5\ ,\qquad
\D_\e  = 1.41264(6)\ ,\qquad
\D_{\e'} =  3.8303(18)\ .
\label{DeltaValues}
\ee
These have been determined by a variety of methods,\footnote{See \cite{cond-mat/0012164} for a review.} like direct experimental measurements, Monte-Carlo simulation, high-temperature expansions,  $\e$-expansion 
and, more recently,  conformal bootstrap techniques.
%
%

We normalize the CFT operators imposing   the following correlation functions in the infinite system without boundaries
\be
\langle \mathcal{O}_i(x)\mathcal{O}_j(y) \rangle = \frac{\delta_{ij}}{(x-y)^{2\D_{i}}}
\ ,\ \ \ \ \ \ \ \ \ \ \ \ \ \ \ \ \
\langle \mathcal{O}(x) \rangle = 0\ , \ \ 
\forall \ \mathcal{O}\neq I \ .
\label{norma2pt}
\ee
This can be used to fix the coefficients in (\ref{ssigma}) and (\ref{eIeps}). The recent Monte Carlo simulations of \cite{arXiv:1501.04065} found
\be
b_{s\s}=0.550(4)\ ,\qquad
b_{eI}=0.330213(12)\ ,\qquad
b_{e\e}=0.237(3)\ .
\label{latticenumbers}
\ee

\section{CFT inside a sphere}

Let us consider the three dimensional critical Ising model in a ball with free boundary conditions on the two dimensional spherical boundary. We would like to test if, in the continuum limit, this system is described by a Boundary Conformal Field Theory (BCFT).  
In order to do this, we start by discussing what are the implications of conformal symmetry for correlation functions in this geometry.

Consider first the one-point function 
$\langle \mathcal{O}(r)\rangle $ of a scalar  operator placed at distance $r$ from the centre of a sphere of radius $R$.
Without conformal invariance, this one point function would take the general form
\be
\langle \mathcal{O}(r)\rangle  =
\frac{1}{R^\D} f\left(\frac{r}{R}\right)
\label{1ptgeneral}
\ee
for an arbitrary function $f$.
Imposing conformal symmetry leads to
\be
f(z)=\frac{a_\Ocal }{(1-z^2)^{\D}}\ ,
\label{1ptconformal}
\ee
for some constant $a_\Ocal$.

In order to see how this comes about, we define the ball 
\be
(x^1)^2+(x^2)^2+(x^3)^2 < R^2\ , 
\qquad
x^\mu \in \mathbb{R}^3\ ,
\label{ball}
\ee 
and introduce new coordinates $y^\mu$ via
\footnote{
It might be helpful to understand this change of coordinates as a sequence of simpler steps. 
Start by doing a translation to bring the point $x^\mu=r^\mu$ to the origin and then perform an inversion. More precisely, defining $w^\mu=(R^2-r^2)(x^\mu-r^\mu)/(x^\mu-r^\mu)^2$ maps the ball $x^2<R^2$ to the exterior of a sphere of radius $R$ centred at $w^\mu=r^\mu$. Moreover, the point 
$x^\mu=r^\mu$ is mapped to $w^\mu=\infty$.
Then, we just translate the sphere in the $w$ coordinates to the origin and perform another inversion, 
$w^\mu-r^\mu=R^2 y^\mu/y^2$.
}
\be  
x^{\mu}  =(R^2-r^2)
\frac{R^2y^\mu +y^2 r^\mu}{ R^4+2R^2y\cdot r+r^2 y^2} +r^\mu \ ,
\label{xintermsofy}
\ee 
where $r^\mu$ is a constant vector with norm smaller than $R$.
This coordinate transformation has several nice properties. The first is that it preserves the spherical boundary. In other words, the ball (\ref{ball}) corresponds exactly to the same region in the $y$-coordinates
\be
(y^1)^2+(y^2)^2+(y^3)^2 < R^2\ ,\qquad
y^\mu \in \mathbb{R}^3\ .
\label{yball}
\ee
The second nice property is that the origin in the $y$-coordinates is mapped to the  point
 $x^\mu=r^\mu$ inside the sphere, which we are free to choose.
The third nice property is that
 the original flat metric becomes
\be
ds^2= dx^\mu dx_\mu = 
\frac{R^4(R^2-r^2)^2}{(R^4+2R^2y\cdot r+r^2 y^2)^2}dy^\mu dy_\mu \equiv
\Omega^2(y)dy^\mu dy_\mu\ ,
\ee
in the $y$-coordinates, \emph{i.e.} it is a conformal transformation.

Correlation functions inside a sphere with a flat metric are equal to correlation functions inside the same sphere but with metric $ds^2 =
\Omega^2(y)dy^\mu dy_\mu$,
\be
\langle \mathcal{O}_1(x_1) \dots 
\mathcal{O}_n(x_n) \rangle_{dx^\mu dx_\mu }
=
\langle \mathcal{O}_1(y_1) \dots 
\mathcal{O}_n(y_n) \rangle_{\Omega^2(y)dy^\mu dy_\mu }\ .
\label{coord}
\ee
Notice that this is true in any theory because it follows just from a relabelling of points without changing the physical geometry.
Remarkably, correlation functions of scalar primary operators
in CFTs also satisfy  
\be
\langle \mathcal{O}_1(y_1) \dots 
\mathcal{O}_n(y_n) \rangle_{\Omega^2(y)dy^\mu dy_\mu }=
\Omega^{-\D_1}(y_1)\dots
\Omega^{-\D_n}(y_n)
\langle \mathcal{O}_1(y_1) \dots 
\mathcal{O}_n(y_n) \rangle_{ dy^\mu dy_\mu }\ .
\label{Weyl}
\ee
In other words, CFT correlation functions transform in a simple way under   Weyl transformations (or local rescalings) of the metric.
Equations (\ref{coord}) and (\ref{Weyl}) together lead to \footnote{From now on, we drop the subscript indicating the metric when it is the standard flat Cartesian metric.}
\be
\langle \mathcal{O}_1(x_1) \dots 
\mathcal{O}_n(x_n) \rangle =
\Omega^{-\D_1}(y_1)\dots
\Omega^{-\D_n}(y_n)
\langle \mathcal{O}_1(y_1) \dots 
\mathcal{O}_n(y_n) \rangle \ ,
\ee
for any parameter $r^\mu$ in (\ref{xintermsofy}).
In the  case of the one-point function, 
one finds
\be
\langle \mathcal{O}(x=r) \rangle =
\Omega^{-\D}(0)\langle \mathcal{O}(y=0) \rangle=
\left(\frac{R^2}{R^2-r^2}\right)^\D
\frac{a_\Ocal}{R^\D}\ ,
\ee
as anticipated in (\ref{1ptconformal}).


For a connected two-point function inside the sphere, choosing $r=x_1$, we obtain
\be
\langle \mathcal{O}(x_1) \mathcal{O}(x_2)\rangle_c \equiv
\langle \mathcal{O}(x_1) \mathcal{O}(x_2)\rangle-
\langle \mathcal{O}(x_1)\rangle \langle\mathcal{O}(x_2)\rangle  
 =
\left[\Omega (0)\Omega(y)\right]^{-\D}\langle \mathcal{O}(0)\mathcal{O}(y) \rangle_c
\ee
where
\be
y^\mu=R^2 \frac{ 
(R^2-x_1^2)x_2^\mu-(R^2-2 \,x_1\cdot x_2+x_2^2)x_1^\mu
}{
R^4-2R^2 \,x_1\cdot x_2+x_1^2x_2^2
}\ .
\ee 
In this way we relate a generic two point function inside the sphere to a two point function where one of the points is at the centre of the sphere.
From spherical and scaling symmetry, it follows  that
\be
\langle \mathcal{O}(0)\mathcal{O}(y) \rangle_c
=\frac{f(\zeta)}{(R^2-y^2)^{\D}}\ ,\ \ \ \ \ \ \ \ 
\ \ \ \ 
\zeta=\frac{R^2-y^2}{y^2}\ .
\ee
Therefore, we conclude that  the  two point function of scalar primary operators inside a sphere is given by
\be
\langle \mathcal{O}(x_1) \mathcal{O}(x_2)\rangle_c =
\frac{R^{2\D}}{(R^{2}-x_1^2)^\D (R^2-x_2^2)^\D}
f_{\Ocal\Ocal}(\zeta)
\label{2ptconformal}
\ee
where 
\be
\zeta= \frac{(R^{2}-x_1^2) (R^2-x_2^2)}
{R^2 (x_1-x_2)^2}\ .
\label{zeta}
\ee

%

The $\zeta \to \infty$ limit corresponds to the two points approaching each other ($x_1 \to x_2$) and it is controlled by the same singularity as the two point function (\ref{norma2pt}) of the infinite system. This gives
\be
f_{\Ocal\Ocal}(\zeta) \approx \zeta^{\D} \ , 
\qquad \qquad \zeta \to \infty\ .
\label{zetainfty}
\ee
The $\zeta \to 0$ limit corresponds to one point approaching the spherical boundary.
This limit is controlled the boundary Operator Product Expansion (OPE)   \cite{Cardy:1984bb, cond-mat/9505127, cond-mat/9610143, arXiv:1210.4258}
\be
\Ocal(z,\vec{z}) = \frac{a_{\Ocal}}{(2z )^\D}
+ 
\frac{a_{\Ocal \tilde{\Ocal}}}{(2z )^{\D-\tilde{\D}}} \tilde{\Ocal}(\vec{z}\,)
+\dots
\ee
where we considered a flat boundary, and used coordinates $\vec{z}$ along the boundary  and the distance to the boundary $z$. Normalizing the boundary operators $\tilde{\Ocal}$ to have unit two point function, $\langle \tilde{\Ocal}(\vec{z}_1)\tilde{\Ocal}(\vec{z}_2) \rangle = |\vec{z}_1-\vec{z}_2|^{-2\tilde{\D}}$, we obtain
\be
f_{\Ocal\Ocal}(\zeta) \approx  a_{\Ocal \tilde{\Ocal}}^2 \,\zeta^{ \tilde{\D}} \ , 
\qquad \qquad \zeta \to 0\ ,
\label{zeta0}
\ee
where $\tilde{\Ocal}$ is the boundary operator with lowest dimension that appears in the boundary OPE of $\Ocal$ (excluding the identity).

We will consider the Ising model with free boundary conditions which is known to be described by a BCFT usually called the \emph{ordinary transition} \cite{Cardy:1984bb, Cardy:1996xt}. This BCFT can be defined by the property that it only has one relevant boundary operator $\tilde{\s}$.
This operator is $\mathbb{Z}_2$ odd and has
$\D_{\tilde{\sigma}}\approx 1.27$  \cite{cond-mat/9804083, Deng:2005dh, PhysRevB.83.134425, arXiv:1502.07217}.
The recent conformal bootstrap study \cite{arXiv:1502.07217} also obtained
\footnote{To estimate $a_\e$ we used the value of the bulk OPE coefficient $\l_{\s\s\e}$ reported in \cite{arXiv:1403.4545}.}
\be
\D_{\tilde{\sigma}}= 1.276(2)\ ,\qquad
a_{\s \tilde{\s}}^2 =0.755(13)\ , \qquad
a_\e=- 0.751(4)  \ .
\label{as}
\ee
The lowest dimension boundary operator in the $\mathbb{Z}_2$ even sector is the displacement operator \cite{Cardy}. The displacement operator $\tilde{D}$ has protected dimension $\D_{\tilde{D}}=3$ and its correlation functions obey Ward identities. In particular,
\be
\int d\vec{w} \langle \tilde{D}(\vec{w}) 
\Ocal(z,\vec{z})\rangle= -\frac{\partial}{\partial z}
\langle  
\Ocal(z,\vec{z})\rangle\ .
\label{Ward}
\ee
The two point function of a bulk and a boundary operator is fixed by conformal symmetry. In particular,
\begin{equation}
 \langle \tilde{D}(\vec{w}) 
\e(z,\vec{z})\rangle=  a_{\e\tilde{D}} 
C_{\tilde{D}}
\frac{ (2z)^{3-\D_\e}}
{\left[ z^2+(\vec{z}-\vec{w})^2\right]^3}\ ,
\label{2ptbulkbound}
\end{equation}
where $C_{\tilde{D}}$ is the normalization
of the two-point function of the displacement operator
\be
\langle \tilde{D}(\vec{w}) 
\tilde{D}(\vec{z}) \rangle = \frac{C_{\tilde{D}}}{
\left|\vec{z}-\vec{w}\right|^3}\ .
\ee
Using (\ref{2ptbulkbound}) and the Ward identity (\ref{Ward}),
we conclude that $4\pi a_{\e\tilde{D}} C_{\tilde{D}}= \D_\e a_\e $.  Finally, we conclude that 
\be
f_{\e\e}(\zeta) \approx 
 C_{\tilde{D}}  a_{\e\tilde{D}}^2
  \,\zeta^{3} 
= \frac{1}{C_{\tilde{D}}}\left(\frac{\D_\e a_{\e}}{4\pi}\right)^2 \,\zeta^{3} \ , 
\qquad \qquad \zeta \to 0\ .
\label{zeta0e}
\ee

\section{Results from Monte-Carlo simulation}

\begin{figure}
\begin{centering}
\includegraphics[scale=0.5]{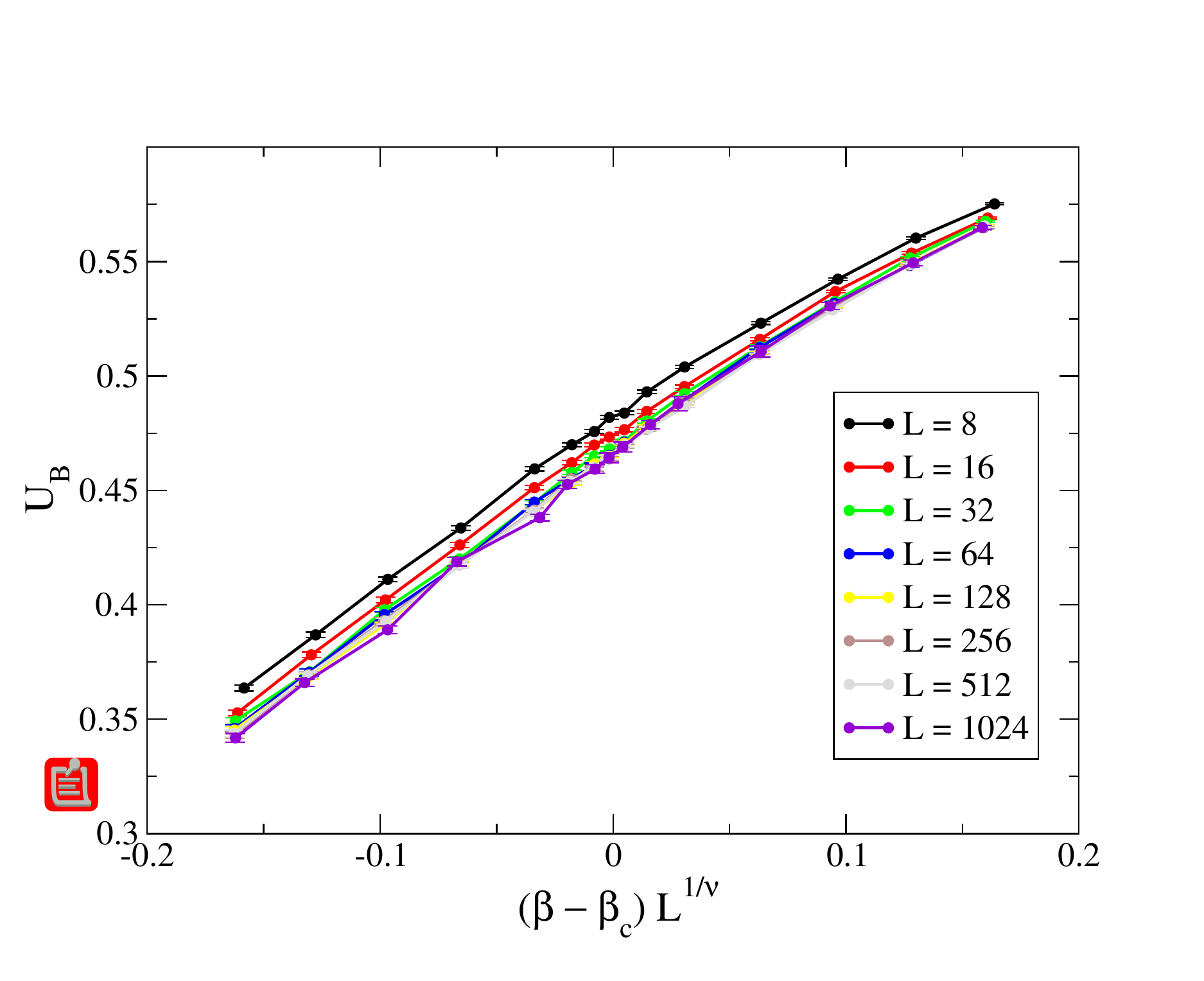}
\par\end{centering}
\caption{\label{fig:Binder}
Binder cumulant for several system sizes.
We used $\beta_c=0.22165463$ and $1/\nu=3-\D_\e=1.58736$. The good collapse of the curves shows that $\b_c$ is sufficiently precise for the large systems we simulated.
}
\end{figure}

In order to perform a Monte-Carlo simulation of the critical Ising model, we need to know the critical temperature with high precision. \footnote{We need to be sufficient close to the critical temperature so that the correlation length is much larger that the size of the system.}
We used  Wolff's cluster algorithm  \cite{Wolff:1988uh} to avoid critical slowing down and used the value 
\be
\beta_c=0.22165463(8)\ .
\label{betac}
\ee
of the critical temperature from 
\cite{PhysRevB.82.174433}.
To check that this is a good estimate of the critical temperature we measured the Binder cumulant 
\be
U_B=1-\frac{ \langle m^4 \rangle}{3\langle m^2 \rangle^2}\ ,
\ee
where $m=\frac{1}{N}\sum_x s(x)$ is the magnetization per spin, with $N=(L/a)^3$ the total number of spins in a   system with  periodic boundary conditions.
In figure \ref{fig:Binder}, we plot the Binder cumulant   for several system sizes.

\begin{figure}
\begin{centering}
\includegraphics[scale=0.3]{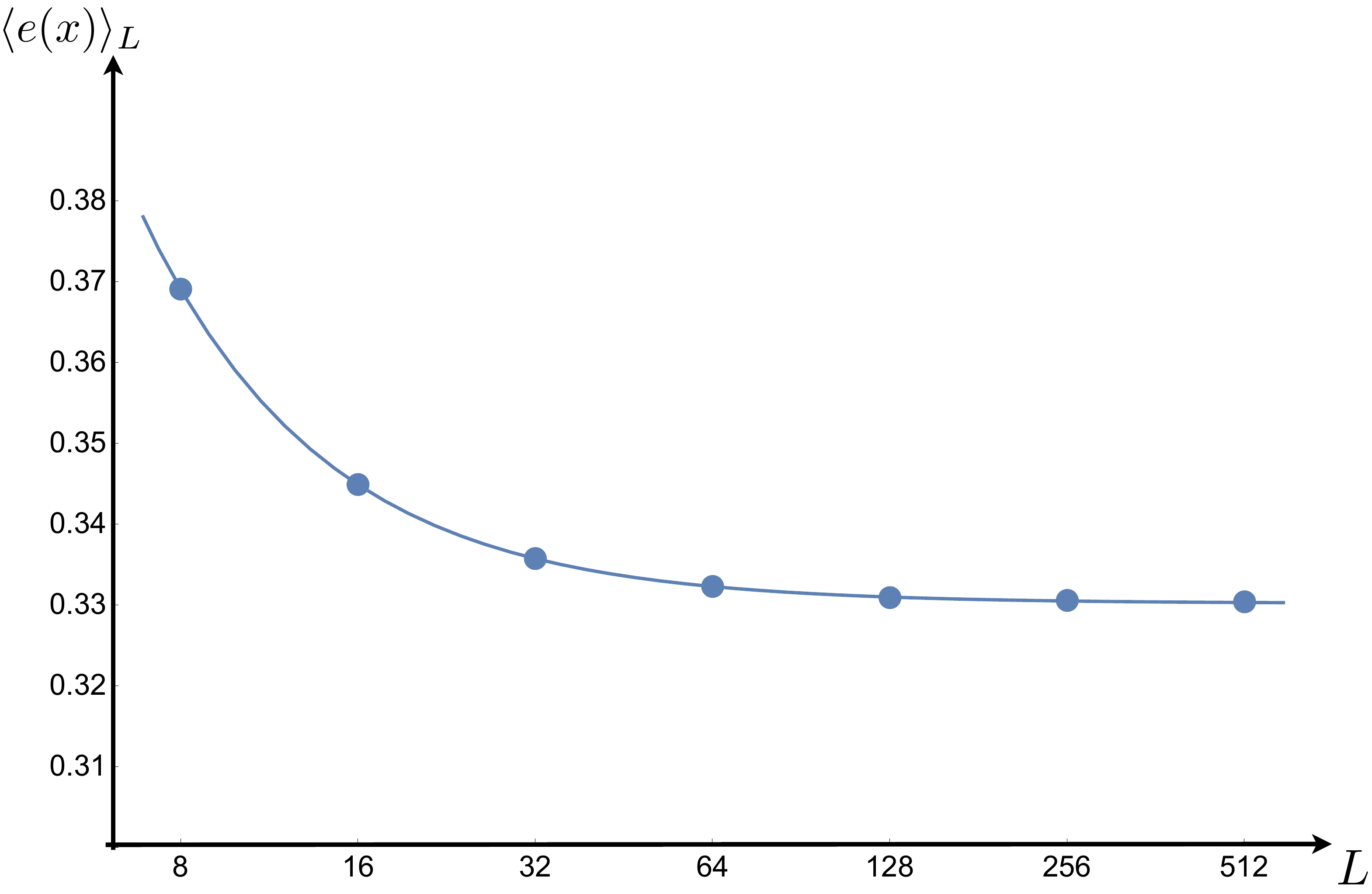}
\par\end{centering}
\caption{\label{fig:eofL}
Expectation values of the energy density operator $e(x)$ at the critical temperature of the 3D Ising model with periodic boundary conditions for several linear sizes $L$ of the system.
The fitting curve is given by the 3-parameter function in equation (\ref{fitEofL})
using the values (\ref{DeltaValues}) for $\D_\e$ and $\D_{\e'}$. 
}
\end{figure}

We also used these simulations with periodic boundary conditions to determine the expectation value $\langle e(x) \rangle=b_{eI}$ of the energy density operator in the infinite system at $\b=\b_c$.
In figure \ref{fig:eofL}, we show the  measured values of $\langle e(x) \rangle_L$ for several system sizes and fit them to the theoretical expectation from equation (\ref{eIeps}) \footnote{The correction proportional to $d$ comes from the leading irrelevant operator in the effective action description of the lattice model.}
\be
\langle e(x) \rangle_L = b_{eI} 
+ c_\e \left(\frac{a}{L}\right)^{\D_\e}\left[1
+ d
\left(\frac{a}{L}\right)^{\D_{\e'}-3}
 +\dots\right] +\dots
 \label{fitEofL}
\ee
Using the values (\ref{DeltaValues}) for $\D_\e$ and $\D_{\e'}$, the fit gave $ b_{eI} = 0.330200(3)$, $c_\e =0.7440(4)$ and $d= -0.084(3)$, in agreement with (\ref{latticenumbers}) and \cite{PhysRevB.85.174421}.

\begin{figure}
\begin{centering}
\includegraphics[scale=0.5]{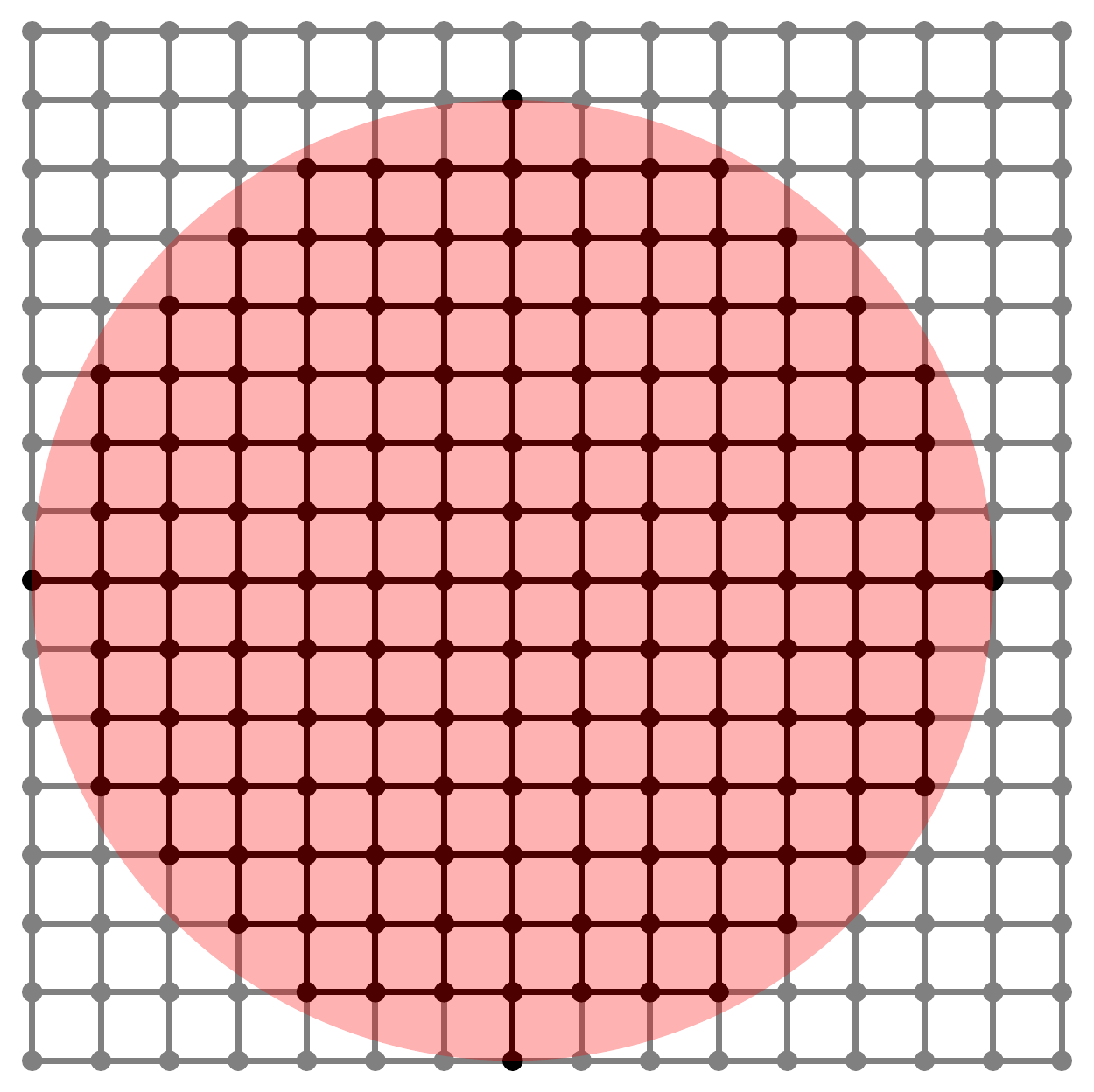}
\par\end{centering}
\caption{\label{fig:LatticeSphere}
Two dimensional cut of a sphere of radius $R=7a$ embedded in a cubic lattice of linear size $L=16$. All spins that are outside the sphere are dropped together with their interaction bounds (in light grey).
}
\end{figure}

We are now ready to compare our results from the Monte-Carlo simulation with the predictions from conformal invariance.
We consider the critical Ising model (\ref{IsingZ})  in a three dimensional cubic lattice excluding all spins outside a sphere of radius $R$ as shown in figure \ref{fig:LatticeSphere}. The interaction bounds connecting spins inside the sphere with spins outside the sphere are also dropped.

\begin{figure}
\begin{centering}
\includegraphics[scale=0.4]{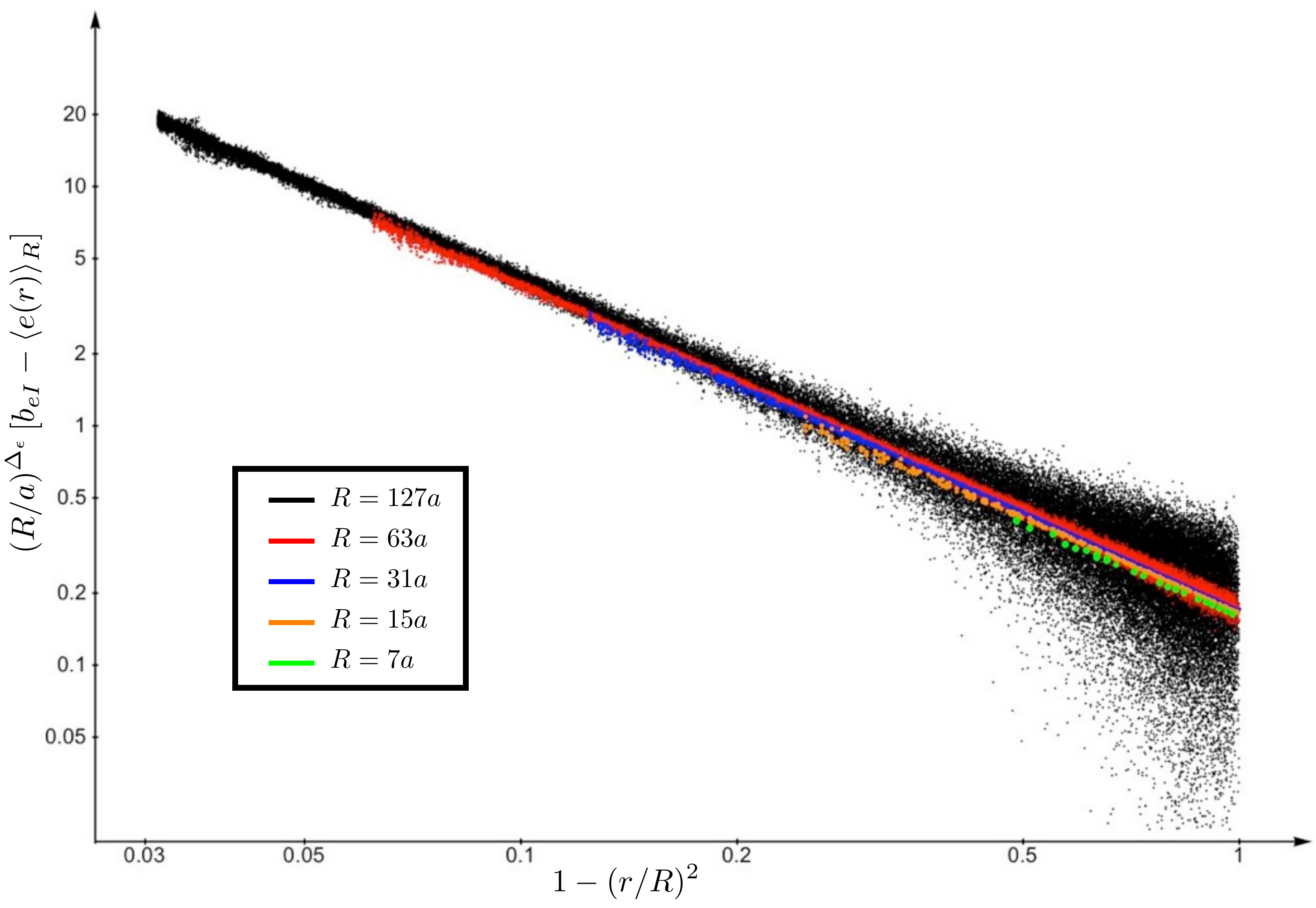}
\par\end{centering}
\caption{\label{fig:1pointsphere}
Expectation value of the energy density operator (minus its value in the infinite system) for several values of the radial coordinate $r$ and of the sphere radius $R$.
The collapse of all points into a single straight line confirms the prediction
(\ref{1pointconformal}) of conformal symmetry.
There are deviations due to finite size effects  and due to statistical uncertainty, specially in the larger systems. 
}
\end{figure}

\begin{figure}
\begin{centering}
\includegraphics[scale=0.4]{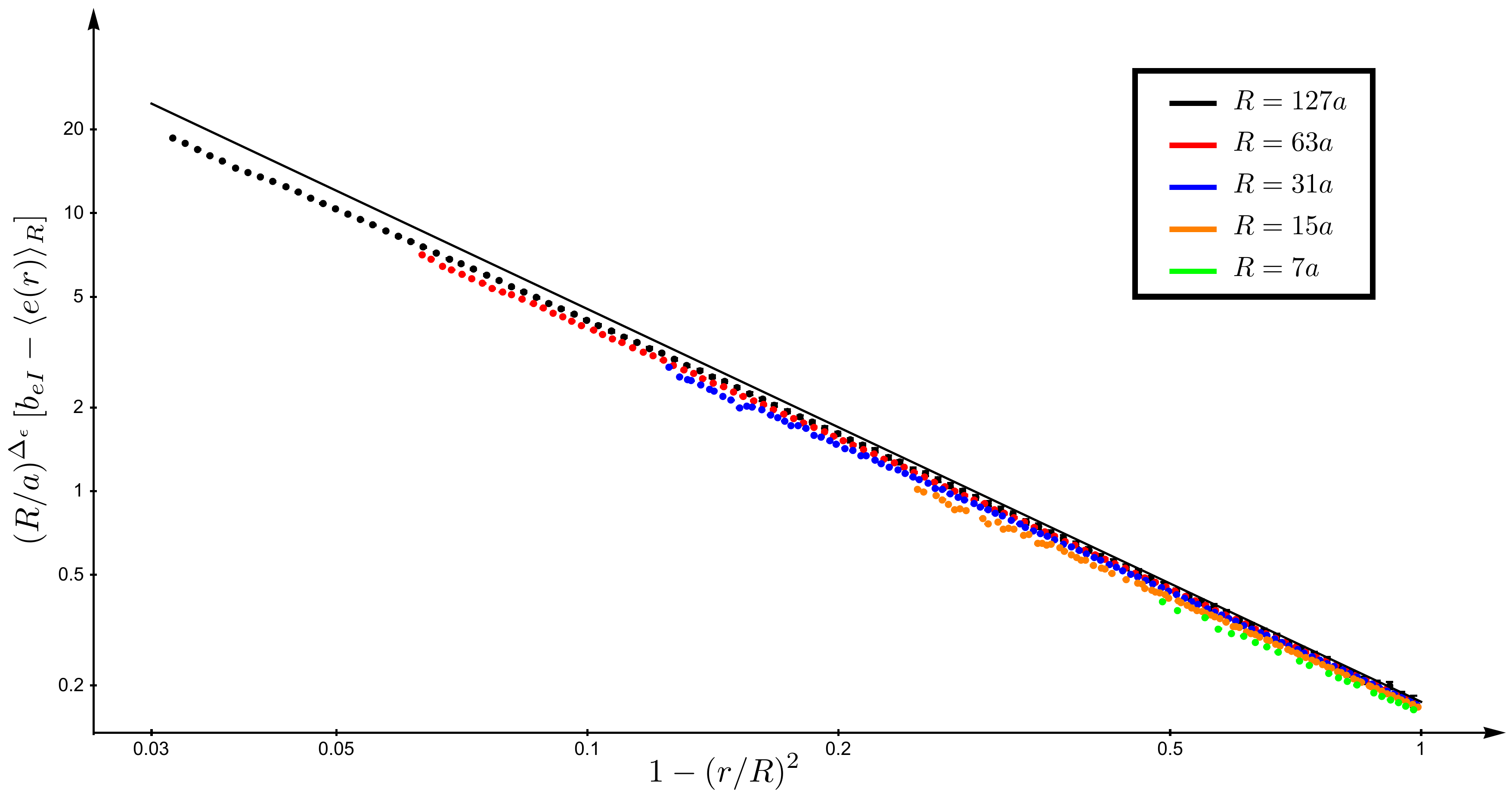}
\par\end{centering}
\caption{\label{fig:1pointsphereBinned}
The same as in figure \ref{fig:1pointsphere} but with the points grouped into 100 bins with approximately the same value of $r$.
The statistical error bars are smaller than the size of the dots.
The black straight line is given by equation
(\ref{1pointconformal}) with
$\D_\e$ from (\ref{DeltaValues}) and 
$b_{e\e}$ from (\ref{latticenumbers}) and $a_\e$ from (\ref{as}).
}
\end{figure}

In figure \ref{fig:1pointsphere} we show the one-point function of the energy density operator inside the sphere, for various radii $R=7a, 15a, 31a, 63a, 127a$.
Combining (\ref{eIeps}) with the prediction (\ref{1ptgeneral}-\ref{1ptconformal}) from conformal invariance, we conclude that
\be 
\left(R/a\right)^{\D_\e}
\big[
\langle e(r) \rangle_R
-b_{eI} \big] =
\frac{b_{e\e}\,a_\e}{\left(1-r^2/R^2\right)^{\D_\e}}+\dots
\label{1pointconformal}
\ee
where the dots stand for terms that vanish in the continuum limit $a/R \to 0$ with $r/R$ fixed.
The plot in figure \ref{fig:1pointsphereBinned} confirms this prediction and the values of $b_{e\e}$ and $a_\e$ given in \ref{latticenumbers}  and \ref{as}.
In figures \ref{fig:1pointsphere} and \ref{fig:1pointsphereBinned} one can notice deviations from spherical symmetry due to the underlying cubic lattice, specially for points close to the spherical boundary.
For large $R$ we have bigger statistical error 
due to the smaller number of independent samples harvested and because the correlation function is multiplied by a large number $(R/a)^{\D_\e}$.
This is our first direct verification of a non-trivial prediction of conformal invariance.
Our second test is related to the two-point function inside the sphere.

\begin{figure}
\begin{centering}
\includegraphics[scale=0.4]{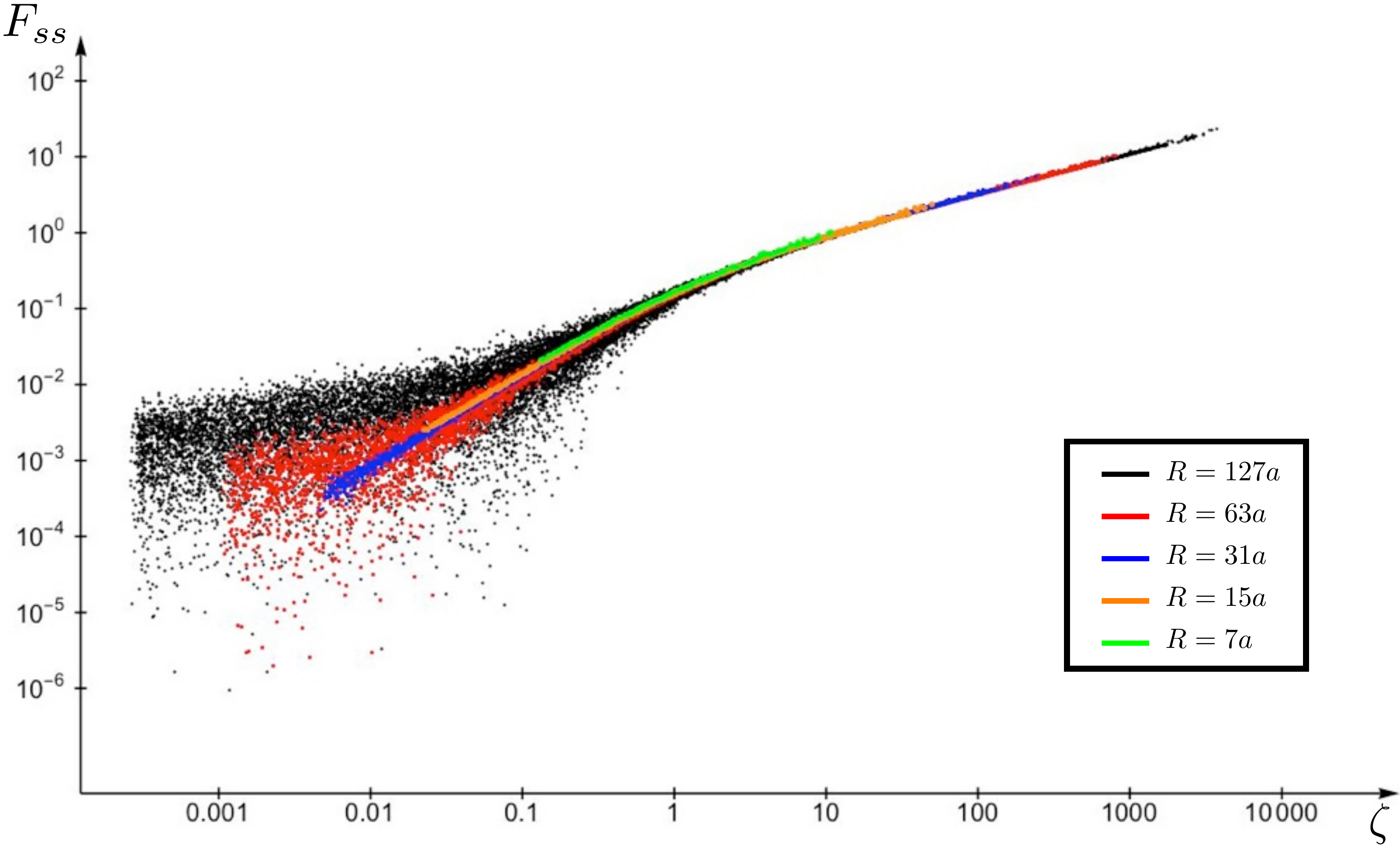}
\par\end{centering}
\caption{\label{fig:2ptspheress}
The combination (\ref{def:Fss}) involving the spin-spin two-point function inside a sphere against the conformal invariant ratio $\zeta$ defined in (\ref{zeta}). Conformal invariance predicts that all points should fall into a single curve up to statistical uncertainties and finite size effects.
}
\end{figure}

\begin{figure}
\begin{centering}
\includegraphics[scale=0.4]{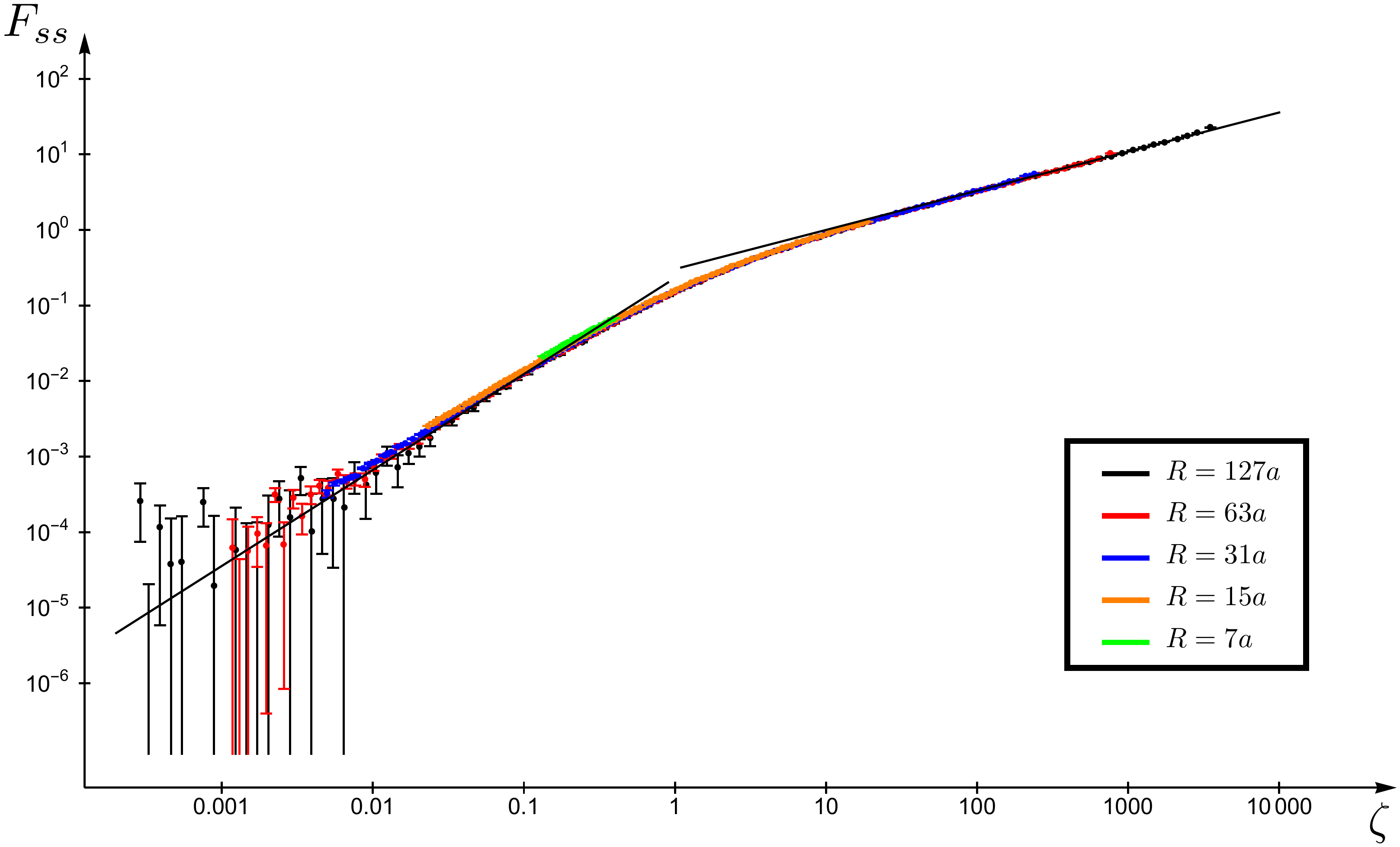}
\par\end{centering}
\caption{
\label{fig:2ptspheressBinned}
The same as in figure \ref{fig:2ptspheress} but with the points grouped into 100 bins for each system size. This reduces the statistical errors and shows better convergence to a single curve. The two straight lines are fits to the asymptotic behaviour using the values (\ref{DeltaValues}), (\ref{latticenumbers}) and (\ref{as}).
}
\end{figure}

From (\ref{ssigma}) and (\ref{2ptconformal}), we obtain
\be
\left(R/a\right)^{2\D_\s}
\langle s(x_1) s(x_2)\rangle_R 
=
\frac{b_{s\s}^2 f_{\s\s}(\zeta)}{(1-x_1^2/R^{2})^{ \D_\s}(1-x_2^2/R^{2})^{ \D_\s}}
+\dots
\ee
where $\z$ is the conformal invariant ratio introduced in (\ref{zeta}) and we neglect terms that vanish in the continuum limit.
In other words, conformal invariance predicts that
the dimensionless function
\be  
F_{ss}(x_1,x_2) =  
 \frac{R^{2\D_\s}}{a^{2\D_\s}} 
\frac{ 
\langle s(x_1) s(x_2)\rangle_R
}{(1-x_1^2/R^{2})^{- \D_\s}(1-x_2^2/R^{2})^{- \D_\s}} 
\label{def:Fss}
\ee
only depends on $x_1$ and $x_2$ trough the combination $\zeta$.
In figures \ref{fig:2ptspheress} and \ref{fig:2ptspheressBinned}, we plot $F_{ss}$ against $\z$ for many different choices of $x_1$ and $x_2$ and for several sphere radii.
As expected, the points collapse in a single smooth curve up to the statistical error bars and finite system size effects.
Moreover, using 
\be  
\lim_{a \to 0} 
F_{ss}(x_1,x_2)
=
b_{s\s}^2 f_{\s\s}(\zeta)
\ee
and the results (\ref{zetainfty}) and (\ref{zeta0}) for the asymptotic behaviour of $f(\z)$
together with the values (\ref{DeltaValues}), (\ref{latticenumbers}) and (\ref{as}), we can verify that $f_{\s\s}(\zeta)$ has the expected asymptotic bahaviour.

\begin{figure}
\begin{centering}
\includegraphics[scale=0.35]{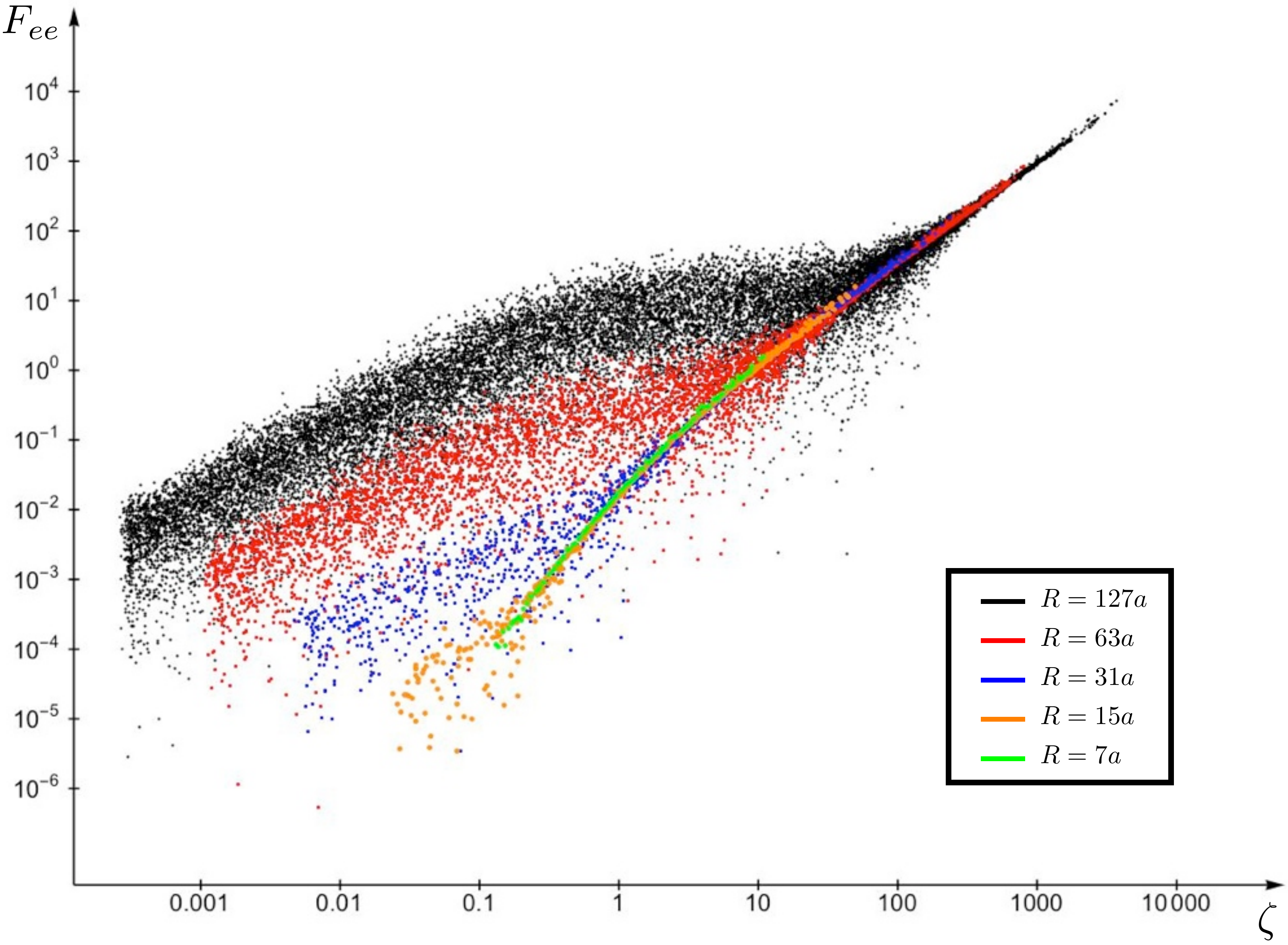}
\par\end{centering}
\caption{\label{fig:2ptsphereee}
The combination (\ref{def:Fee}) involving the  connected two-point function of the energy density operator inside a sphere against the conformal invariant ratio $\zeta$ defined in (\ref{zeta}). Conformal invariance predicts that all points should fall into a single curve up to statistical uncertainties and finite size effects.
The statistical uncertainty looks biased because we are using a logarithmic scale and therefore  we can not plot the points with $F_{ee}< 0$.
}
\end{figure}

\begin{figure}
\begin{centering}
\includegraphics[scale=0.35]{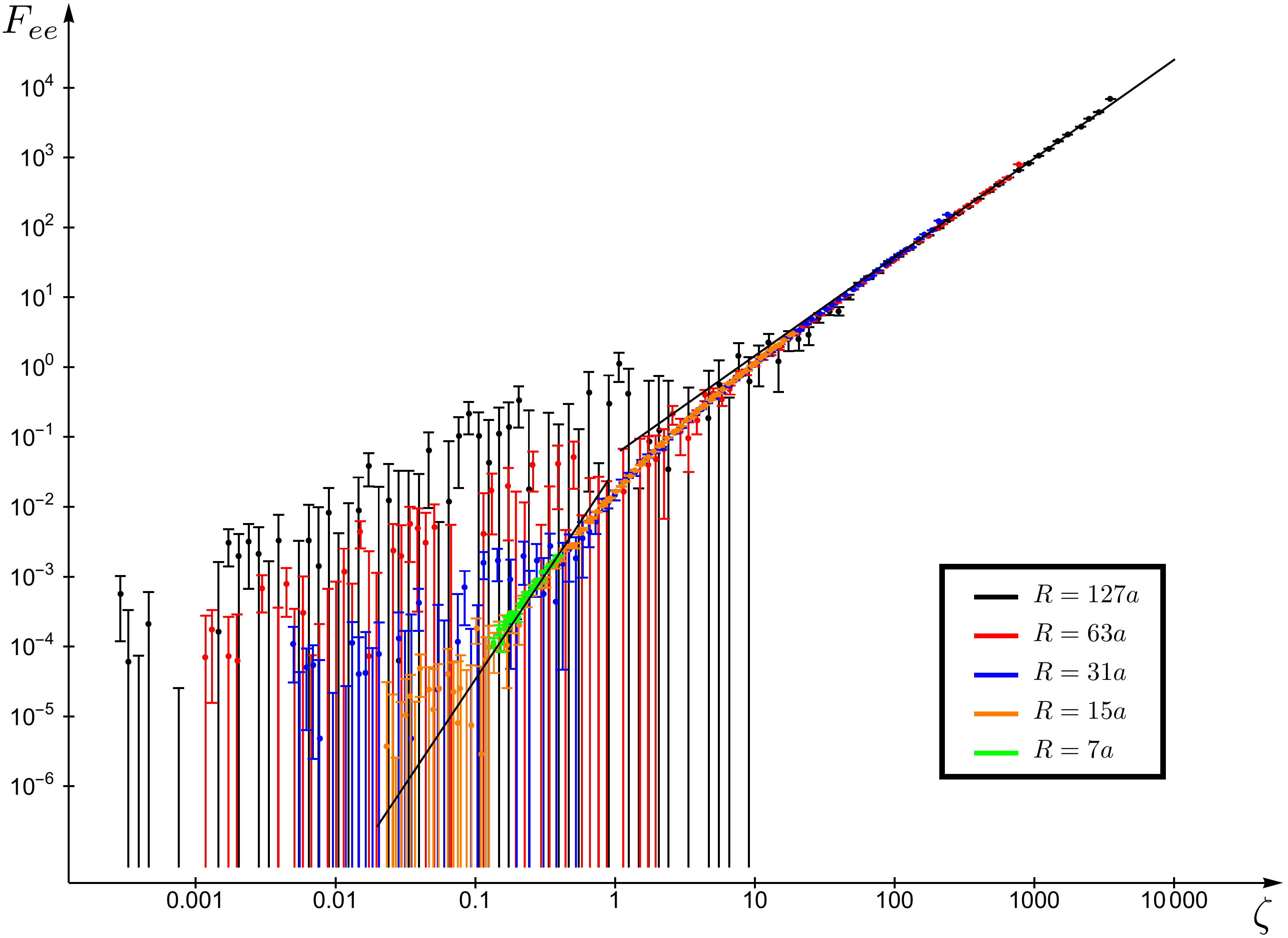}
\par\end{centering}
\caption{
\label{fig:2ptsphereeeBinned}
The same as in figure \ref{fig:2ptsphereee} but with the points grouped into 100 bins for each system size. This reduces the statistical errors and shows better convergence to a single curve, although the errors are still large for small values of $\zeta$. The two straight lines are  the expected asymptotic behaviour using the values (\ref{DeltaValues}), (\ref{latticenumbers}), (\ref{as}) and $C_{\tilde{D}}= 0.012$.
}
\end{figure}

We also preformed a similar analysis for the two point function of the energy density operator inside the sphere. 
In figures \ref{fig:2ptsphereee} and \ref{fig:2ptsphereeeBinned}, we plot
\be  
F_{ee}(x_1,x_2) =  
 \frac{R^{2\D_\e}}{a^{2\D_\e}} 
\frac{ 
\langle e(x_1) e(x_2)\rangle_R -
\langle e(x_1)\rangle_R \langle e(x_2)\rangle_R  }{(1-x_1^2/R^{2})^{- \D_\e}(1-x_2^2/R^{2})^{- \D_\e}} 
\label{def:Fee}
\ee
against the conformal invariant $\zeta$.
Using the bulk scaling dimensions \ref{DeltaValues} and the fact that the lowest dimension $\mathbb{Z}_2$-even surface operator is the displacement operator with $\D_{\tilde{D}}=3$, we can fit  the asymptotic behaviour of the curve to conclude that $C_{\tilde{D}}\approx 0.012$.
The last  value is a very crude estimate  because $F_{ee}$ has very large statistical uncertainty in the region of small $\zeta$. The main reason for this is that in this region the correlation function is very small and it takes a long time to simulate the large systems required to explore the $\zeta \to 0$ limit.


\section{Conclusion}

We gave strong evidence confirming the 
 non-trivial predictions of conformal symmetry for correlation functions of the critical Ising model in a ball geometry.
We hope our work strengths the confidence in the conformal bootstrap methods that assume conformal symmetry from the start.

It would be nice to obtain more precise measurements of scaling dimensions and OPE coefficients of several boundary operators. However, the cubic lattice discretization of the ball geometry we used is not ideal for this purpose because it introduces large finite size and boundary effects.
It would also be interesting to study other BCFT of the critical Ising model, like the \emph{special} and the \emph{extraordinary} transition. 
These can be implemented introducing another coupling between the boundary spins. 
Unfortunately, it is not obvious how to do this in an elegant fashion in our ball geometry.

In the absence of boundaries, a fundamental prediction of conformal symmetry is the functional form of three-point correlation functions.
We plan to verify this prediction with Monte-Carlo simulations, in the same spirit of this paper.
Such study would also be able to check several  conformal bootstrap predictions for OPE coefficients of the Ising CFT.
It is curious that in two dimensions this was done 20 years ago \cite{Barkema:1995mw}.

\section*{Acknowledgements}
We are grateful to Slava Rychkov for useful discussions and for suggesting this work.
The research leading to these results has received funding from the [European Union] Seventh Framework Programme [FP7-People-2010-IRSES] and [FP7/2007-2013] under grant agreements No 269217, 317089 and No 247252, and from the grant CERN/FP/123599/2011. 
\emph{Centro de Física do Porto} is partially funded by the Foundation for  Science and Technology of Portugal (FCT). 
J.V.P.L. acknowledges funding from projecto Operacional Regional do Norte, within Quadro de Referência Estratégico Nacional (QREN) and through Fundo Europeu de Desenvolvimento Regional (FEDER), Ref. NORTE-07-0124-FEDER- 000037.

\bibliographystyle{./utphys}
\bibliography{./bibIsing}

\end{document}